# Atomic-Scale Investigation on the Ultra-large Bending Behaviours of Layered Sodium Titanate Nanowires

*Qiong Liu, Haifei Zhan, Huaiyong Zhu, Ziqi Sun, John Bell, Arixin Bo\* and Yuantong Gu\**

School of Chemistry, Physics and Mechanical Engineering, Queensland University of Technology, GPO Box 2434, 4001, Brisbane, QLD, Australia.

**Abstract**

Study on mechanical properties of one-dimensional layered titanate nanomaterials is crucial since they demonstrate important applications in various fields. Here, we conducted *ex situ* and *in situ* atomic-scale investigation on bending properties of a kind of ceramic layered titanate ($Na_2Ti_2O_4(OH)_2$) nanowires in a transmission electron microscopy. The nanowires showed flexibility along <100> direction and could obtain a maximum bending strain of nearly 37%. By analysing the defect behaviours, the unique bending properties of this ceramic material was found to correlate with a novel arrangement of dislocations, an accessible nucleation and movement along the axial direction resulting from the weak electrostatic interaction between the $TiO_6$ layers and the low *b*/*a* ratio. These results provide pioneering and key understanding on bending behaviours of layered titanate nanowire families and potentially other one-dimensional nanomaterials with layered crystalline structures.

**Introduction**

One-dimensional (1D) nanomaterials have received extensive research interest in recent years along with the vast development of microelectronics,[1] energy,[2] environment,[3] biology,[4] and other fields. Similar to the applications of bulk materials, nanomaterials in integration and practical applications are subjected to mechanical deformation, especially in flexible devices and single nanowire (NW) based electronics.[1,5,6] Also, the intriguing physical properties of NWs can be exploited when they are under mechanical strain.[7] Therefore, besides the physical and chemical properties of NWs, it is important to understand their mechanical behaviour. However, it is challenging to extract the mechanical properties of NWs due to the limited accessibility of mechanical characterization at the nanoscale.

In recent decades, investigations on the mechanical characteristics of nanostructured materials have been an active topic.[8-12] Previously, the majority of the research on the mechanical properties of the NWs mainly focused on the strength and Young's modulus.[8,9] It is well-known that mechanical properties of a material are determined by the atomic configuration and bonding types within the structure. Metals generally show ductile behaviour while ceramics are usually stiff and brittle.[11,12] Meanwhile, the defect behaviour in the materials plays an important role in the mechanical properties. Therefore, it is necessary to clarify the structural evolution and the effects of defects on the mechanical properties of nanomaterials, which usually requires the use of *in situ* atomic-scale experimental methods.[10] In the bulk, ceramics are usually brittle at room temperature with a small strain at fracture of $0.1 - 0.2\%$.[13] Such a small strain results from the lack of the dislocation slip caused by the high Peierls-Nabarro potential, which leads to the fast formation and propagation of cracks.[12,14] However, large bending strain and plasticity can be observed when the dimensions of ceramics are reduced to the nano/micro scale, which can be found in Si,[15,16] ZnO,[17] SiC,[13] and WC[14] NWs. The phenomena are derived from the scarcity of defects as well as the nucleation and movement of dislocations.

Layered inorganic nanomaterials, formed by stacks of nanosheets, have attracted great attention due to their unique structure and applications in Na-ion batteries,[18,19] bioscaffolds,[20] water treatment,[21] and photocatalysts.[22] Among this family, layered titanate nanomaterials are composed of repeating sheets of $TiO_6$ octahedrons with exchangeable cations accommodated in between the layers. This kind of unique crystalline structure endows them with novel physicochemical properties and thus renders them important and widely studied in various fields, such as photocatalysis,[22] sensors,[23] batteries,[19,24] and water treatment.[25] The continuous development of efficient 1D energy storage device as well as promotion of sodium ion batteries makes the investigation on mechanical deformation an important priority.[26,27] Previous research has used different kinds of methods to study the mechanical properties of nanomaterials, like using a bimetallic-strip device[10,28] or a microelectromechanical system (MEMS) device[11] as the test stage and using an atomic force microscopy (AFM) probe as a manipulator in scanning electron microscopy (SEM) or transmission electron microscopy (TEM).[12,14,17] Recently, Bo *et al* investigated the mechanical properties of layered titanate NWs where the Young's modulus and yield strength of $Na_2Ti_3O_7$ NWs were measured by using AFM.[29] However, to the best of our knowledge, *in situ* bending deformation of layered titanate NWs has not been carried out and the underlying mechanisms for their bending properties in room temperature remain unknown. Thus, it is significant to offer experimental observation on the crystalline structure evolution of layered titanate NWs during the bending deformation. Apart from the ionic and metallic bonds in the $TiO_6$ layers, there are also electrostatic interactions between the interlayer cations and the negatively charged layers. The role of interlayer cations is to stabilise the layered structures by reducing the electrostatic repulsion between the negatively charged layers. The structural complexity and the variety of atomic interactions can lead to extreme mechanical anisotropy. For instance, the layers in 2D layered titanates can undergo slippage when a shear force is applied to the planes, which offer the potential for use as a solid lubricant.[30] Thus, it can be speculated that intriguing mechanical performance may be observed during mechanical deformation for the layered titanate NWs.

In this work, we investigated the bending properties of $Na_2Ti_2O_4(OH)_2$ (NTO) NWs which were grown on Ti foil via a facile hydrothermal method. The NTO has a body-centered orthorhombic layered structure with $a$ = 1.926 nm, $b$ = 0.378 nm, and $c$ = 0.300 nm).[31] *In situ* bending tests for the NWs were conducted by using the colloidal-film-induced bending method inside a TEM.[10,13] The method can protect the NWs from contamination and potential structural damage that materials may experience during sample preparation processes, where techniques, like focused ion beam, are required to transfer and fix the materials onto the AFM probe or testing device.[14,29] The NWs showed flexibility in <100> direction which is perpendicular to the planes of $TiO_6$ layers with a comparative large bending strain of ~6% generated accessibly by the surface tension of ethanol dispersant. Fracture occurred at a local bending strain of 37% when further imposing bending deformation on the NW using a colloidal-film-induced bending method.[13,33] Through *ex situ* and *in situ* observations on the NWs during deformation processes by high-resolution TEM (HRTEM), the dislocation nucleation and movement were observed, as well as a series connection of dislocations in the layered crystalline structure of the NTO NWs, which accounts for their large bending strains.

**Materials and methods**

*Materials*

Thin NTO NWs were synthesized by a one-step hydrothermal method.[32] Typically, a piece of titanium foil (2 × 1.5 cm$^2$) was ultrasonically cleaned in acetone, ethanol, and deionized water for 20 min respectively. Then, the Ti foil was placed in a 120 mL Teflon-lined autoclave filled with 60 mL 1 M NaOH solution. The autoclave was maintained at 220 °C for 14 h and cooled down to room temperature naturally. Finally, the white coloured Ti foil was taken out and dried at 80 °C for 12 h after being rinsed in deionized water for several times.

*Sample characterization*

The morphologies and the microstructure of the samples were observed by a field emission SEM (JEOL JSM-7001F) and a TEM (JEOL 2100, 200 kV). The crystal structure was characterized by X-ray diffraction (XRD) using a Philips PANalytical X'pert pro diffractometer equipped with a graphite monochromator with Cu Kα radiation of 40 KV. HRTEM images and selected area electron diffraction (SAED) patterns were taken by using the JEOL 2100 TEM to analyze the microstructure of the material. The elemental composition was defined by energy-dispersive X-ray spectroscopy (EDS) that is attached to the TEM.

*TEM sample preparation*

To characterize the NTO NWs in TEM, the NWs were scratched from the Ti foil and dispersed in ethanol by ultrasonication. Then the mixture was dropped onto a holey carbon-film-supported TEM copper grid that was placed on a piece of filter paper. It should be noted that the concentration of the solution was carefully controlled to avoid the aggregation of multiple NWs. The NW that sits across holes of the carbon film will be probably bent towards the edge of the carbon film by the surface tension of the ethanol with its gradual evaporation.

*In situ bending test in TEM*

The *in situ* bending test was conducted by irradiating the supporting carbon film attached on the TEM copper grid using the focused electron beam.[13,33] An individual NW lying on the edge of a hole of the carbon film was selected as the testing sample. The thin carbon film shrinks under the irradiation of the electron beam, resulting in a tensile force laterally applied on the NTO NW to bend it. This *in situ* TEM technique can allow investigating the mechanical behaviours at the atomic scale.

**Results and discussion**

*Structural characterization*

The XRD spectra were collected to define the crystal structure of the material, as shown in Fig. 1a. Apart from the diffraction peaks from Ti substrate, the other peaks index well with the (200), (110), (310), (020), and (220) of $H_2Ti_2O_4(OH)_2$ (JCPDS, 47-0124), which is identical to NTO.[31,34] As shown in the low-magnification SEM image (Fig. 1b), the entire surface of the Ti foil was covered by a dense vertically grown NW array with uniform cross sections. The average length of the NWs is about 5 μm. The top-view image (Fig. 1c) shows a circular cross section of the NWs with the diameter ranging from 20 to 70 nm.

The morphological and crystalline structures of the NWs were further characterized using TEM. A low-magnified TEM image (Fig. 1d) shows that the diameter is uniform for each NW. The EDS results show that the NW contains Ti, O, and Na (Fig. S1). The HRTEM image was taken from the framed area in Fig. 1d, showing a single crystalline feature of the NWs. Two sets of lattice fringes are marked, which are parallel and normal to the growth direction (i.e., <010> direction), respectively. Their inter-planar spacings are 0.204 nm and 0.193 nm, which can be attributed to $d_{800}$ and $d_{020}$, respectively. However, the lattice spacing of 0.204 nm of (800) is smaller than the theoretical value of 0.240 nm. This may be derived from the dehydration of the NW, leading to the shrinkage of the lattice in <100> direction.[35] The inset SAED pattern that is viewed along the [001] direction in Fig. 1e shows a single set of diffraction spots, which further confirms the single crystalline nature of the NWs.

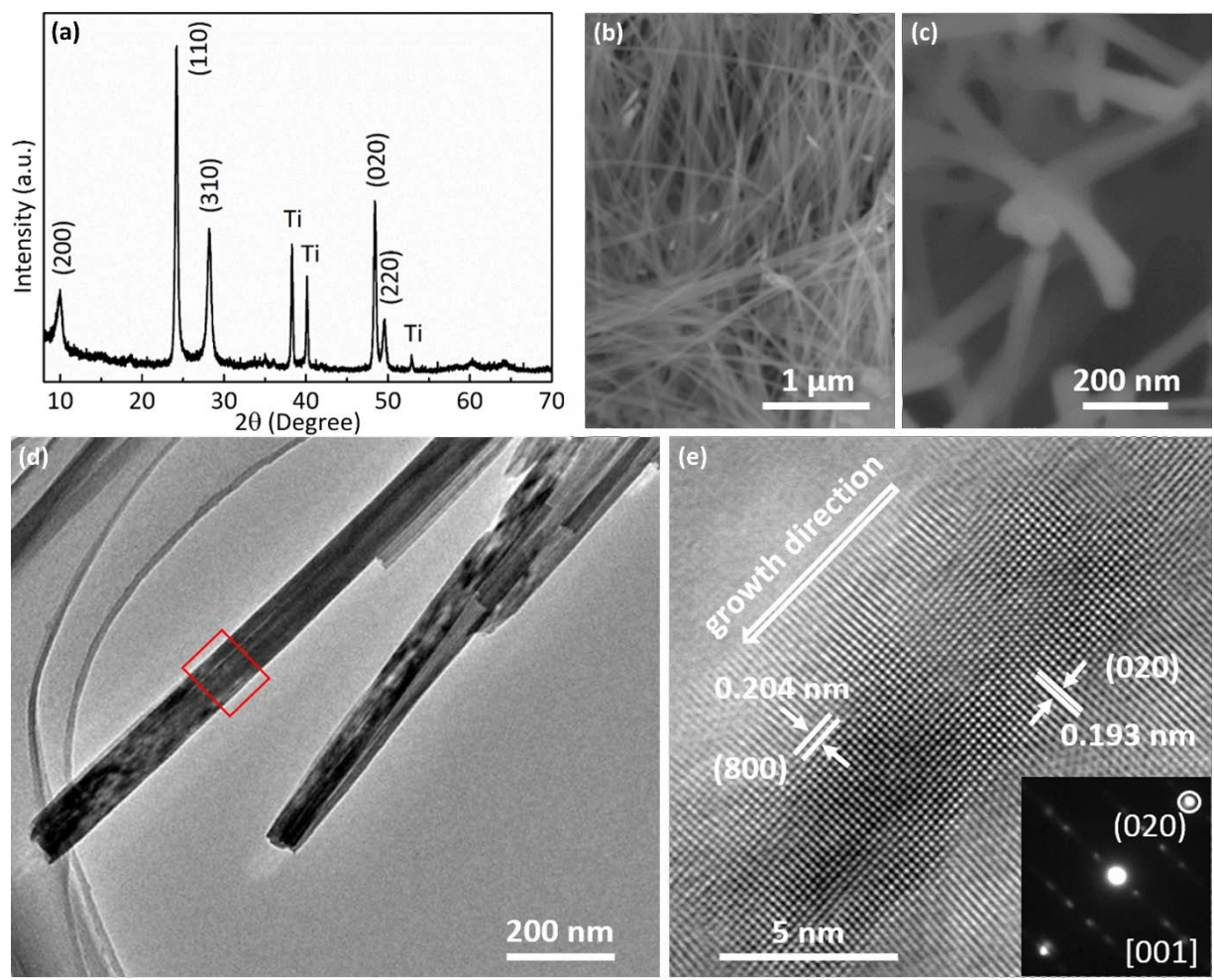

**Fig. 1** (a) XRD pattern of the as-grown NTO NWs. (b) low-magnified SEM image, (c) higher-magnified SEM image, and (d) low-magnified TEM image of the NTO NWs. (e) HRTEM image of the framed area in (d) with a SAED pattern being inset.

Fig. 2a shows another typical low-magnified TEM images of the NTO NWs. As is shown, a bent NW lies along the curvature of the carbon film. The bent NWs that were located near the carbon film could be found at multiple places on the TEM grid, like another individual bent NW that is shown in Fig. 2b. The bending deformation is believed to be the result of ethanol-induced surface tension, deforming the NW as the ethanol evaporates during the TEM sample preparation. At room temperature, evaporation of the ethanol led the NW beam lying across the hole of the carbon film to be pulled towards the carbon film, which indicates the flexible nature of the NW. Through this way, an initial bending strain of 5.7% could be obtained, as shown in Fig. 2c. The strain is calculated by $\varepsilon = r/\rho$, where $r$ is the radius and $\rho$ is the bending curvature of the NW, respectively.[13]

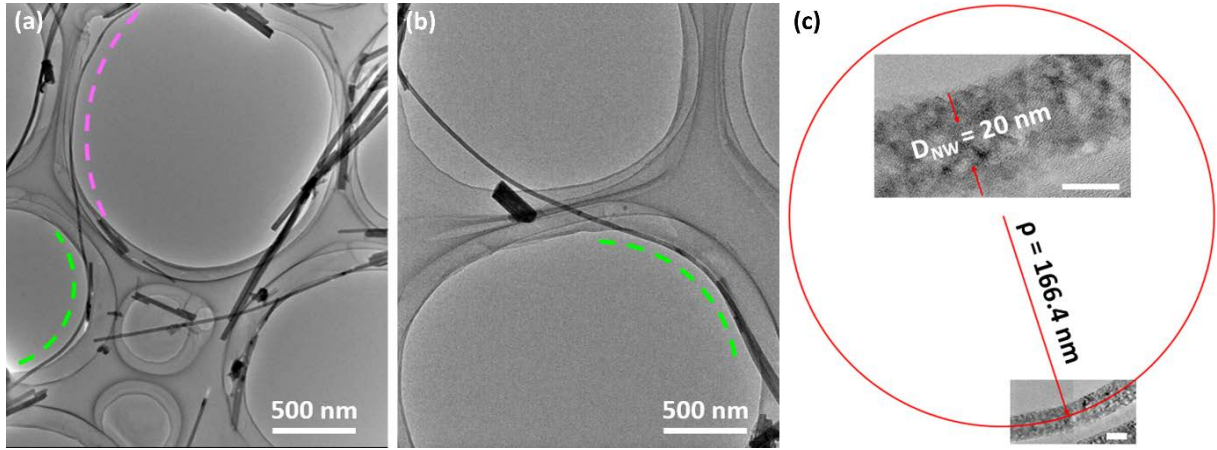

**Fig. 2** (a,b) TEM images of two bent NWs, respectively, after the evaporation of ethanol. (c) The strain calculation of a bent NW. The scale bars in the top image and the below image in (c) are 20 nm.

*Ex situ bending deformation*

The initial bending strain of the NW marked by the green arc in Fig. 2a is 3.0% which is a comparative large value for a ceramic NW. The strain is comparable to the elastic strains of GaN (2.5%),[36] VO$_2$ (3.8%),[37] and SiC (2%) NWs before fracture or amorphization.[13] To better reveal the mechanical performances of NTO NWs, a larger bending strain needs to be achieved. Such operation can be implemented through heating the carbon film to generate a mechanical force inside a TEM chamber. Fig. 3a and b show the same NW marked by the green arc in Fig. 2a before and after the further bending deformation, respectively. After irradiating the thin carbon film for about 30 min by the focused electron beam, the shrinkage of the carbon film deformed the NW, eventually causing the fracture of the NW (framed area in Fig. 3b). Higher magnification over the deformed area shows clear fracture of the NW (inset in Fig. 3b). It is worth mentioning that the beam irradiation was placed on the carbon film instead of on the NW to avoid potential impact of electron beam irradiation. Also, during the observation, the electron beam dose rate was kept at a level of around $3 \times 10^{19}$ e/(cm$^2$·s) which led the local temperature to be only a few degrees higher than the room temperature.[13] Thus, the thermal impact was minimized to ensure that the deformation of the NW was solely derived from mechanical impact. Upon fracture, the local bending strain near the compressive surface is measured to be about 37% which should reach the fracture strain of the NW. This value is considered quite large for ceramic NWs and found to be larger than many other ceramic NWs. For instance, the SiC NWs experienced amorphization with the bending strain above 2%.[13] In Wang's work, WC NWs were completely fractured after getting a bending strain of about 23%.[14] The uncommon ductility during bending of the NTO NWs is most likely related to the dislocation behaviour, as discussed below.

HRTEM images corresponding to the fracture regions extracted from Fig. 3a and b are depicted in Fig. 3c and d, respectively. With a bending strain of 3.0%, numerous dislocations can be seen, which are shown by the red arrows (Fig. 3c). Such a high density of dislocations is found to be within the (020) planes and is mostly likely caused by the bending strain. As Fig. 1e indicated, dislocations can hardly be observed in a strain-free NTO NW. The high density of dislocations resulting from a low bending strain suggests that a low amount of energy is required to activate the formation of dislocations lying in (020) planes. This may account for the flexible nature of the NWs that they tend to bend towards the <100> direction. Moreover, both the compressive and the tensile sides of the NW contain more dislocations than the neutral region in the centre of the NW, which derives from the more intense strain fields in the two sides. The two inset fast Fourier transformation (FFT) patterns in Fig. 3c correspond to the top and bottom regions: region I marked by the blue square and region II marked by the orange square, respectively. By measuring the angle between the two (020) fringes in the two FFT patterns, the size of the bending is determined to be a small angle of 3.2°.

Fig. 3d shows the HRTEM image of the same region in Fig. 3c after a further bending deformation. It is clear that the NW was torn apart at the tensile side, leading to a 14° angle between the (020) fringe in region I and (020) fringe in region II, which is shown in the inset FFT patterns. Meanwhile, the compressive side exhibits a complex lattice pattern. The bending strain is indicated by a 21° angle whose two sides are marked by the blue and orange line, respectively. The large angle bending is compensated by dislocations lying in the (020) planes in the angle. A 7° angle next to the 21° angle is compensated by a (020) dislocation that heads opposite to the direction of dislocations in the 21°. The difference of the two angles is consistent with the 14° angle measured by the FFT patterns above. Moreover, even though cracking occurred at the NW tensile side, (800) planes maintained their continuity under compressive strain as marked by the curved line. This indicates that slip preferentially occurred on the (800) planes where the comparatively weaker electrostatic interactions are found. The weak atomic interactions between the $TiO_6$ layers together with the rather low $b/a$ ratio resulted in a low shear modulus on the (800) planes, resulting in a low Peierls-Nabarro potential.[14] Thus, under the bending strain the dislocation slip was accessible in the brittle NTO NWs, especially favourable in the (800) planes where the $TiO_6$ layers are lying, which can retard the fast propagation of dislocations and the following possible cracks.

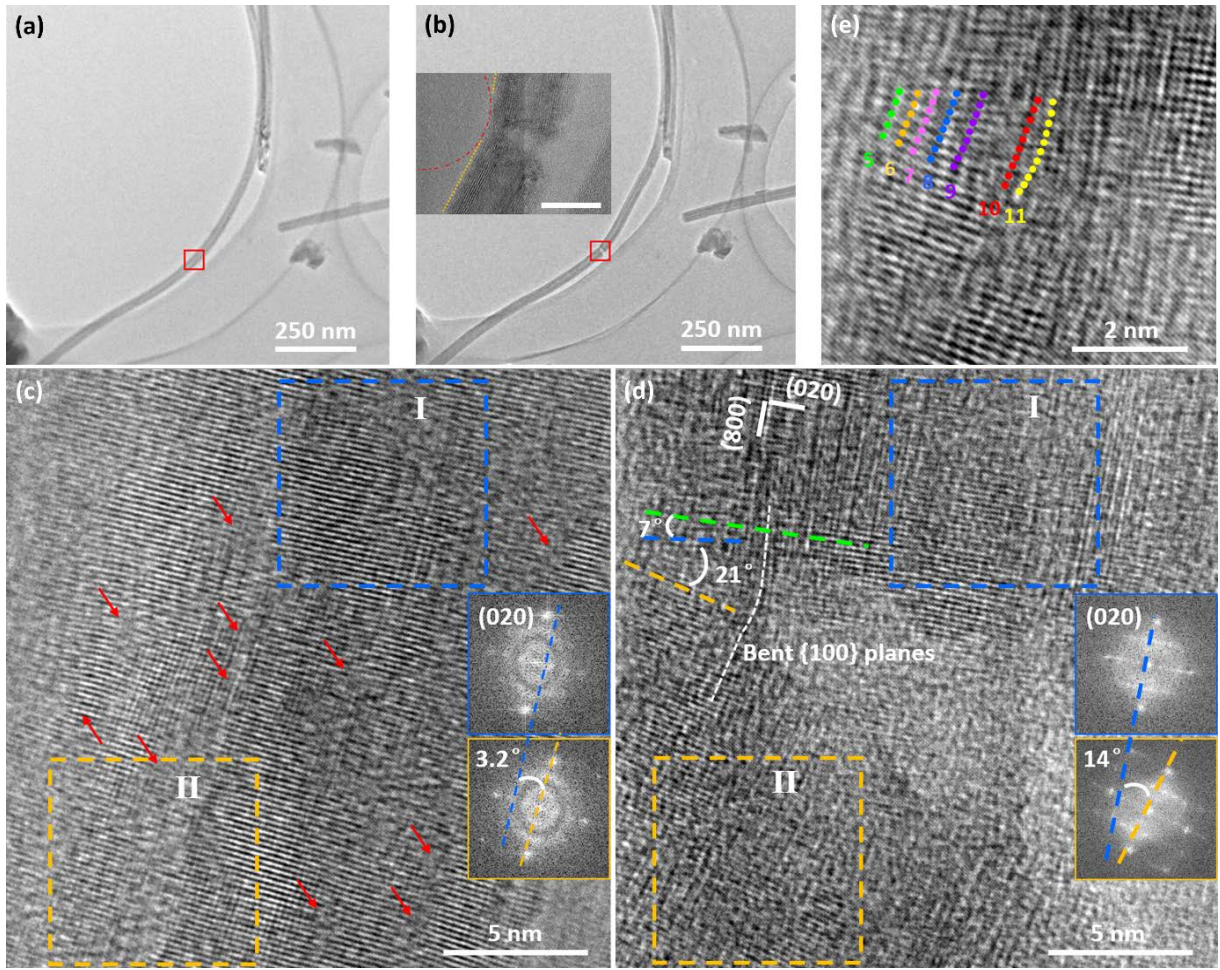

**Fig. 3** Low-magnified TEM images of a bent NW (a) before and (b) after a further bending deformation. Inset is the enlarged TEM image of the framed area in (b) with a scale bar of 10 nm. (c,d) HRTEM images corresponding to the same area framed in (a) and (b), respectively. (e) HRTEM image of the compressive region in (d).

To better understand the ultra-large bending behaviour of the NW, the pattern of dislocation distribution in the projection of the (020) planes in the 21° angle is illustrated in detail in the corresponding HRTEM image (Fig. 3e). It can be observed that the number of (020) planes increases from the compressive surface to the inner part of the NW. Specifically, the (020) layers have been marked out by the dots to reveal the progressive increase of the layer quantities and the dislocation alignment. At the left-most part, five recognizable layers are marked by the five dots and the adjacent right dots number six layers. Thus, a dislocation can be defined in between the two dot rows. Right next to the six-dot row is a seven-layer row, from which another dislocation is found to be situated in between the two dot rows. In the same way, the quantity of the layers increases gradually with the distance getting larger from the compressive surface. Seven positions have been lined out in total, numbering from five layers to eleven layers, which indicates that six <100>-oriented dislocations are in this area. Also, the six dislocations were found to form a connection in series. As we know, under a small strain the bending deformation can be sustained with the emergence of a small quantity of dislocations in ceramic NWs. More dislocations are generated under a large bending strain, which are usually arranged disorderly and propagate quickly, either causing a break-off or amorphization of the NWs.[13,14] However, for the NTO NWs under an ultra-large bending strain, the dislocation pattern of a connection in series was still able to compensate the large bending deformation without causing amorphization or complete break-off of the NWs.

*In situ bending deformation*

Fig. S2b-d illustrate the bending process of an individual NTO NW shown in Fig. S2a. The strain increased from an initial value of 1.2% to 5.9% then to 9.8%. Fig. 4a-c are the HRTEM images captured from the same region marked by the red boxes in Fig. S2b-d, respectively. From the three images, no large-scale disorderly atomic distribution or plastic deformation can be observed. Fig. 4d-f display the lattice patterns of the compressive part in the same area (marked by the red boxes, respectively) of the NW with different bending strains. As can be seen in Fig. 4d, with a minor strain of 1.2%, no dislocations are observed. When the strain reached 5.9%, several full dislocations lying in (020) planes (marked by "⊥") nucleated (Fig. 4e) in this area. With a larger strain of 9.8%, more dislocations can be observed (Fig. 4f), indicating that the bending deformation was accompanied by the dislocation nucleation and movement. Also, it can be observed that lattice distortions occurred in (020) planes (encircled regions). Fig. 4g-i show the lattice patterns of the tensile part in the same region (marked by the orange boxes) with strain increasing. Only one dislocation is observed in the viewed area, initially (Fig. 4g). When the strain increased to 5.9%, the dislocation in Fig 4g disappeared, with a new dislocation appearing in this area (Fig. 4h). Meanwhile, several cases of (020)-plane distortions were observed. When the NW was subjected to a bending strain of 9.8%, a high density of dislocations in the (020) planes and one dislocation in (800) plane were generated (Fig. 4i). This phenomenon further confirmed that the large bending strain was attributed to the dislocation activities.

Typically, it is difficult for dislocation motion to occur in ceramic materials due to the scarcity of the slip systems resulting from the directional ionic and/or covalent bonds. Dislocations tend to propagate and lead to fracture of the materials when they are under a large mechanical strain.[38] When the dimensions of the material are reduced to nanoscale, the number of dislocations may be decreased, which enable ceramic NWs sustain a larger mechanical strain.[39] However, after the dislocation nucleated in a ceramic NW, the dislocation motion will be hindered, resulting the overall structure unable to support a very large strain.[12] Nevertheless, some NWs such as SiC NWs could behave a large bending deformation by experiencing amorphization through dislocation motion.[13] For the NTO NWs, the large bending strain was also related to the dislocation motion. The evolution of lattice structures in both compressive and tensile sides of the NTO NW indicates that the dislocation nucleation and movements happened frequently and more accessibly in the NW during the bending deformation. The active slip system is {100}<010>, which means that the dislocation slip mainly occurs in the longitudinal direction of the NTO NWs. The movements of dislocations can protect them from fast propagation and retard the following crack of the NW even under a large bending strain. Such dislocation motion was also reported in WC NWs with large bending strain. Differently, the large bending strain of the WC NWs was also balanced by the amorphization at the tensile surface.[14]

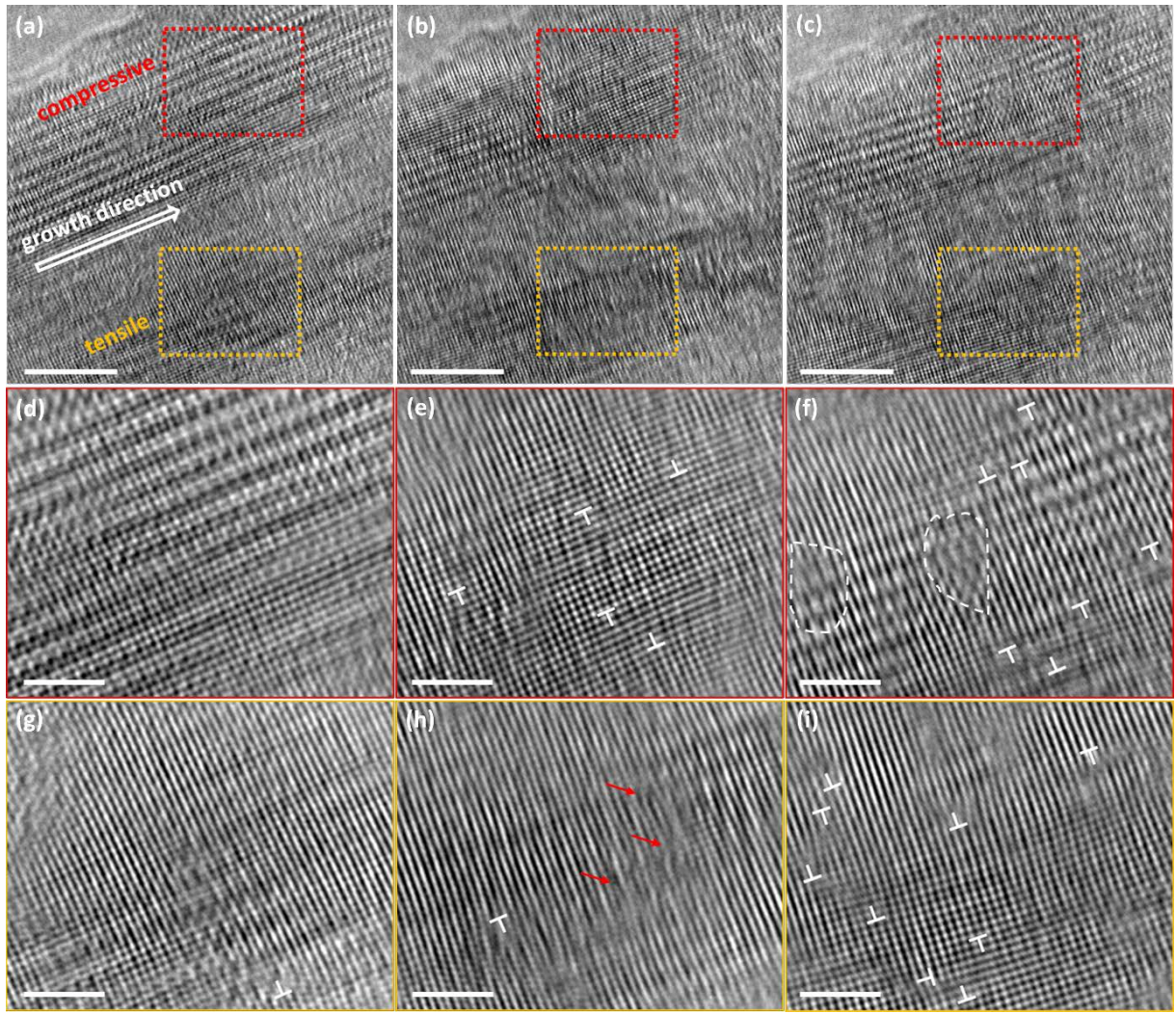

**Fig. 4** *In situ* observation of the bending process of an individual NTO NW. (a-c) HRTEM images of the same region marked by the boxes in Fig. S2b-d with the strain of 1.2%, 5.9, and 9.8%, respectively. (d-f) Enlarged HRTEM images of the same area located at the compressive side of the NW, as marked by the red boxes in (a-c), respectively. (g-i) Enlarged HRTEM images of the same area located at the tensile side of the NW, as marked by the orange boxes in (a-c), respectively. The scale bars in (a-c) are 5 nm, and (d-i) are 2 nm.

*Dislocation motion*

To confirm the feasibility of the dislocation motion especially the slip of dislocations in (020) planes along the {100} family of lattice planes, a minor force disturbance was applied on a NW by heating the nearby carbon film after it was fractured. The low magnified TEM images are shown in Fig. S3a and b. The mechanical deformation of the NW was imperceptible under low magnification. Therefore, the neighbouring carbon film and the apex of the broken NW (pointed out by the two arrows, respectively) imaged at high-magnification were as the references to monitor the movements of the NW, as shown in Fig. S3c-f. It can be seen that the apex of the NW was reaching out to the carbon film gradually. After broken off, the NW was unable to recover because of the attachment between the carbon film and the NW. The whole process is most likely to reflect a relaxation of the residual strain near the end of the NW.

Fig. 5a-d are the enlarged HRTEM images that correspond to the same region marked by the boxes in Fig. S3c-f, respectively. The *in situ* observation reveals the dislocation nucleation and movements during the slight bending deformations of the NW. Fig. 5a shows the initial lattice pattern of this area immediately after the NW was broken. Three dislocations have been highlighted by Burgers circuits among which dislocation #1 has a Burgers vector of $b/2[0\bar{1}0]$, and dislocations #2 and #3 have Burgers vectors of $a/8[100]$ and $a/8[\bar{1}00]$, respectively. During NW moving towards the carbon film slightly, dislocation #1 kept stationary while dislocations #2 and #3 disappeared from the viewed area and a new dislocation #4 nucleated with a Burgers vector of $b/2[010]$, as shown in Fig. 5b. The vanishing of dislocation #2 and #3 might be due to their colliding with each other. No dislocations lying in the (800) planes were nucleated again in the viewed area during the next two deformation stages, as shown in Fig. 5c and d. The phenomena indicate that it requires more energy to activate the dislocation nucleation in (800) planes than that in (020) planes. With further deformation, dislocation #4 disappeared and a new dislocation #5 nucleated which is also lying in the (020) planes while dislocation #1 slipped across several layers along the (800) plane (marked as 1'), as shown in Fig. 5c. Continuing pulling the end of the NW, a lattice distortion was generated in the (020) planes, as marked by the arrow. Moreover, either dislocation #1' or #5 slipped to a different location (marked as 1'' and 5', respectively) while the distance of the two dislocation cores maintained unchangeable. This further confirms that the lattice slip prefers to occur along (800) planes, which results from the comparative weak electrostatic interactions between the $TiO_6$ layers and the low $b/a$ ratio. The activated dislocation movements helped these dislocations avoid propagation, which was beneficial to disperse the bending energy in the NW, increasing the fracture strain of the NTO NWs. This the first direct observation of defect behaviours for bent layered titanate NWs. The deformation mechanisms related to the layered crystalline structure for the NTO NWs may be applicable to other layered ceramic NWs.

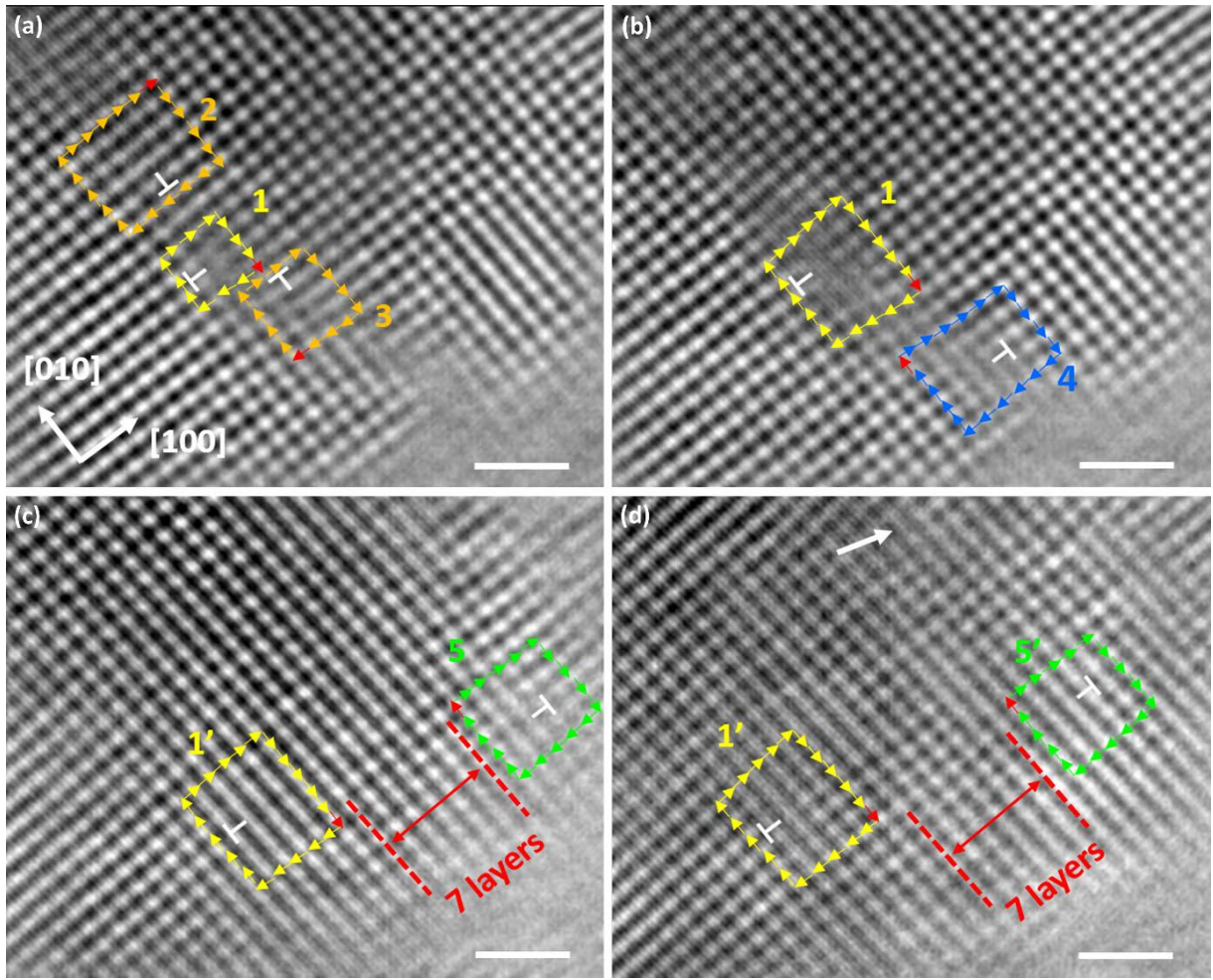

**Fig. 5** *In situ* observation of the defect nucleation and movements under a minor mechanical deformation. (a-d) Enlarged HRTEM images corresponding to the same area marked by the boxes in Fig. S3c-f, respectively.

**Conclusion**

In summary, the bending properties of $Na_2Ti_2O_4(OH)_2$ NWs were investigated. The NWs showed flexibility along <100> direction which is normal to the planes of $TiO_6$ layers. An ultimate strain of 37% for the NWs could be observed. Through *ex situ* and *in situ* atomic-scale study, the bending deformation processes of the NWs were found to involve the accessible nucleation and movements as well as a novel arrangement of dislocations because of the unique layered crystalline structure of this material. Provided that the layered crystalline structure is possessed by many other titanates, the results can endow fundamental understanding of the bending behaviours of the layered titanate NW families.

**Acknowledgements**

We acknowledge the support from the Australian Research Council (ARC) Discovery Project (DP170102861), Central Analytical Research Facility (CARF) of Queensland University of Technology (QUT).